\documentclass[reprint,amsmath,amssymb,aps,prl,notitlepage,twocolumn]{revtex4-2}

\usepackage{graphicx}% Include figure files
\usepackage{dcolumn}% Align table columns on decimal point
\usepackage{bm}

\newcommand{\x}{{\bf r}}
\newcommand{\K}{{\bf k}}

\begin{document}

\author{Pawe{\l} Zin}

\affiliation{National Centre for Nuclear Research, ul. Pasteura 7, PL-02-093 Warsaw, Poland}
%\email{pawel.zin@ncbj.gov.pl}

\title{Proof the non-existence of causal classical electrodynamics of point charged particles}

\begin{abstract}
Up until now, a consistent causal theory of point charged particles (for example electrons) interacting with electromagnetic field is not known. The well-known problem is that the standard Lorentz force alone (in the case of point particles) does not lead to a theory in which momentum and energy are conserved. The need of radiation reaction force (or self-force) thus arises. The well known candidate for such force, the Lorentz-Abraham-Dirac reaction force, gives non-causal particle behavior, i.e., the particle starts to move before the arrival of external electromagnetic fields. Alternative, causal proposals provide non-physical behavior of the particle -- the particle moves with non-zero acceleration long after any external forces acted on it. Below, we question the existence of a causal theory. We show that for certain electromagnetic pulse of radiation and point particle being initially at rest, there {\it does not exists} a causal particle trajectory, such that the particle ends up moving with constant velocity (when no external electromagnetic fields act on it for very long time). This shows that the proper, causal electrodynamics of point particles does not exist.
\end{abstract}
 
% Inne propozycje daja te rozwiazania samoprzyspieszon.
% Okazuje sie ze wszystkie teorie cazualne dadza samoprzyspieszone
% - nie istnieje kazualna poprawna trajektoria.
% 

\maketitle

It is well known that in the case of point charged (elementary) particles the Lorentz force is not enough to ensure energy and momentum conservation \cite{Jakson,Rohlich}.
Charged  particles subject to acceleration emit electromagnetic radiation, which carries energy and momenta not accounted by Lorentz force. Thus to restore energy and momentum conservation there is need for a force through which a charged particle acts on itself, causing the effects connected with 
energy and momentum change due to radiation.
This force is called self-force or radiation reaction force.

To illustrate the need for such force, consider a positive point charge moving towards a static positive charge  (held still by external forces). 
For simplicity we assume that  velocity of the moving charge points to the center of the static charge.
In this case the external field that acts on the moving charge is the electric field generated by
the static charge. The direction of the electric field is opposite to the velocity of the moving charge. We now take the Lorentz force to be the only force acting on the moving charge.
Then  according to Newton equation, the final velocity of the moving charge shall be the same as the initial velocity but pointing in opposite direction. 
Eventually, when the charges are far away from each other, the kinetic energy of the moving charge shall be equal to its initial kinetic energy.
On the other hand according to Maxwell equations the moving charge as it accelerates (it changes its velocity so it has to accelerate),  emits electromagnetic radiation. This radiation carries energy. Thus at the end of the process the total energy is larger (by the emitted radiation energy) than the initial energy of the system -- note the lack of energy conservation.
This example clearly illustrates the need of an additional force to restore the conservation laws.

Its derivation was a subject of many investigations \cite{Lorentz,Abraham,Dirac,Feynman,Birula,kij0,kij01,kij,Yag1,Yag2,Spohn,Medina,dopiska}.
The consistent formulation of classical electrodynamics of point particles, known to the author, are due to Dirac \cite{Dirac}
and Kijowski \cite{kij0,kij01,kij}.
A crucial problem of the Dirac's theory is that the particle trajectories, obtained from the derived equation of motion,
are, at least in some cases, non-causal \cite{Dirac,Rohlich}.  
The charged particle, initially at rest, starts to move  {\it before} the external fields reach the particle.
This behavior leads to serious problems (discussed in what follows).
On the other hand Kijowski's approach is causal, but it leads to non-physical particle behavior -- the 
charged particle long after any external electromagnetic fields act on it, moves with non-zero acceleration.
Due to this behavior the theory cannot be considered as a proper one. 
There were also other proposals  (see for example \cite{Yag2,Medina}).
However, up to know, none of them succeeded in providing a bona fide theory.

We now move to the formulation of the theory we analyze.
We do not start from the usual Langrangian formulation, which we shall discuss later on.
We follow instead the formulations used by Dirac \cite{Dirac} and  partly by Kijowski \cite{kij}.
We define standard classical electrodynamics by taking the evolution of electromagnetic fields 
as being given by Maxwell equations with sources being point particle. 
The Maxwell equations provided us with the formulas for energy and momentum densities of the electromagnetic fields
(see Supplementary Material for the detailed discussion why this formulas accompany the Maxwell equations). 
These are $\frac{\epsilon_0}{2} ( E^2 + c^2 B^2)$ and $ \epsilon_0 \, {\bf E} \times {\bf B} $ for energy and momentum density respectively \cite{Jakson}.

The fact that we deal with point particles generates serious difficulties. 
To discuss it we consider the charge at rest for infinitely long time.
In such a case the coulomb electric fields generated by the static particle fills the space.
This fields generates a nonzero energy density equal to $  \frac{\epsilon_0}{2}  E^2$.
Summing this density over the whole space besides the ball of radius $r$ around the charge, gives a positive energy that scales like $1/r$. 
Thus, it increases with the decrease of $r$.
Eventually at certain $r$ that we denote as $r_0$ 
the discussed energy reaches the value equal to the total energy of the charged particle $m c^2$ where $m$ is the mass of the particle.
Still there exist positive energy present inside the ball of radius $r_0$.
And this energy is infinite for a point particle. 
This causes a serious problem in  the construction of the theory of charged point particles.
The need of so called ``renormalization" procedure arises -- 
we need a procedure that gives a recipe for how to subtract the infinite energy (and in general momentum) of the electromagnetic field
and in return add the finite energy (and momentum) of the point particle.
In the case in question this renormalization procedure would be simply to subtract the total (infinite) energy of the coulomb electric field
and in return add $mc^2$.

A charged point particle generates electromagnetic fields.
As it is known there are two types of fields generated by the particle  \cite{Jakson,Rohlich}.
First are the fields attached to the particle, e.g., the coulomb electric field
in the case of static particle. This fields are so to say ``attached" to the particle.
The second type of fields is electromagnetic radiation.
This fields so to say ``detach" from the particle - the field survives even if the particle disappears.
When particle moves with constant velocity for infinitely long time we can use the renormalization procedure in the reference frame comoving with the charge.
 In such a frame we deal with particle which is static. Here we can use the renormalization procedure defined above.
Returning back from the comoving frame to the initial one we find that the renormalization procedure subtract the energy and momentum of the attached fields and in return gives $\gamma m c^2$ , $\gamma m {\bf v}$ where $\gamma = (1-v^2/c^2)^{-1/2}$.

In what follows  we consider the following situation.
Initially, we have a static point charge (resting for infinitely long time)
and a external electromagnetic pulse being very far away from the charge.
As  time flows, the pulse reaches the charge, acts on it for finite time and then leaves the charge.
In such a system we use the renormalization procedure twice:
in the ``initial" and ``final" situation -- very long time before and very long time after the pulse acts on the particle \cite{dopiska4}.
The use of renormalization procedure enables us to find ``initial" and``final" energy and momentum of the system.  As we want  the energy and momentum conservation to take place, therefore we assume that the energy/momentum in initial time is equal to the same quantities at final time.

Performing the calculations outlined above (see Supplementary Supplementary Material for details) we derive two formulae describing energy and momentum conservation, respectively. More precisely, these read
\begin{eqnarray}\label{enN}
&&   \int \mbox{d} t \  \, {\bf E} \cdot {\bf v}  =  \gamma_f -1   +    E_{rad}  
\\ \nonumber
\\ \label{momN}
&&   \int \mbox{d} t \  \left({\bf E} + {\bf v} \times {\bf B} \right) =
 {\bf v}_f \gamma_f  +   {\bf P}_{rad}. 
\end{eqnarray}
Note that here  ${\bf E}$ and ${\bf B}$ ( equal to ${\bf E}(\x(t),t)$  and  ${\bf B}(\x(t),t)$ where  $\x(t)$ is the particle position at time $t$)
denote the fields of the {\it external electromagnetic pulse}. 
We also denoted ${\bf v}(t)$ as particle velocity and $q$ as the particles charge.
In addition $E_{rad}  $ and ${\bf P}_{rad} $ denote the energy and momentum of the fields radiated by the particle and are well known \cite{Jakson}.
Here $\gamma_f = 1/\sqrt{1-v_f^2}$ (see Supplementary Material for discussion of the units used here).
%The quantities  $E_{rad}$ and  ${\bf P}_{rad} $ describe energy and momentum of the fields radiated by the particle.
The terms $\gamma_f -1 $ and $ {\bf v}_f \gamma_f $ come from the energy and momentum of the fields attached to the particle
and the use of renormalization procedure described above.

We need to stress out the fact that the above were obtained by integration over the whole space at initial and final time.
At first glance it is surprising that in the above equations we notice  only temporal integrals and not the spatial ones, which we start from.
However this can be understood by realizing that the radiated waves move with speed of light. Therefore when moving towards the particle through the 
space we find fields which were radiated in different times (the closer to the particle the later time it was).
This is in fact the reason the spatial integral effectively changes into temporal one.
For example the terms $ - \int \mbox{d} t \  \, {\bf E} \cdot {\bf v} $
and  $ - \int \mbox{d} t \  \left({\bf E} + {\bf v} \times {\bf B} \right) $
arise from the integration of the energy and momentum density term that comes from {\it interference} between the
field of the pulse and the field generated  by the particle.
We clearly see that they describe the work and change of momentum due to the Lorentz force.
We want to emphasize here that the above equations were derived directly from the form of 
energy and momentum density of electromagnetic fields and not 
assuming any known formula for Lorentz force.
However it is not surprising that we somehow ``derived" the Lorentz force.

Eqs.~(\ref{enN}) and (\ref{momN}) are due to equality of initial and final energy/momentum of the system.
Knowing the external pulse profile we search for particle trajectory $\x(t)$ that
satisifies Eqs.~(\ref{enN}) and (\ref{momN}). 
In principle we can have many of such trajectories.
The theory we search for should give us a single trajectory -- we need to choose one from many.

Kijowski and coworkers make such choice in formulation of their theory.
They introduce renormalization procedure valid at all times.
As a consequence they are able to calculate energy/momentum of the system at any time.
Their causal theory is defined by setting these quantities constant -- equal to its initial values.
Such formulation is able to give unique particle trajectories.
Unfortunately these trajectories are unphysical
-- the particle undergoes nonzero acceleration as time tends to infinity.
Thus even when external forces do not act on the particle for very long time, it still accelerates.

Still Kijowski and coworkers try only two different renormalization procedures valid at all times. 
In principle we could search for renormalization procedure that would give
causal and physical trajectory. If we find such we would have a desired theory.

We follow a different path instead. Instead of searching for 
the desired theory we treat the result of Kijowski and coworkers, not
as an incorrect  choice of renormalization procedure, but rather a permanent property of causal theories \cite{jeszczeDopiska}.
%In what follows we consider only theories  (with certain choice of renormalization procedure
%vaid at all times) in which for all possible external pulses the causal trajectories exist .
We show that in {\it all} causal theories, for certain shape of external pulse (which we show
below), the resulting causal particle trajectories are unphysical -- they have nonzero acceleration and never tend to constant final velocity.

We proof the above {\it add absurdum} (proof by contradiction).  
We first notice that in the desired theory (causal and physical particle trajectory)  the initial and final  energy/momentum of the system are equal and therefore  Eqs.~(\ref{enN}) and (\ref{momN}) are forfilled. 
Now comes the most important result of this paper. We show that for 
certain external pulse the causal and physical trajectory forfilling 
 Eqs.~(\ref{enN}) and (\ref{momN}) {\it does not exists}!
As the particle trajectory exists  (we consider theories in which trajectiories always exists) it has to be unphysical - posses the behaviour described above.
%It means that the causal particle trajectory has to posses the unphysical behavior described above.

However this means that the causal electrodynamics of point particles does not exists!
If it would, the causal physical trajectory would exist for any shape of the external pulse.

%Looking and Kijowski result we may state the above thesis in different way:
%For {\it all} causal theories with the initial conditions considered here (also with the specific choice of external pulse), the trajectory given by the theory shall have non-physical behavour -- nonzero acceleration for time tending to infinity.
%Thus the result of Kijowski and coworkers is not due the incorect 

We now show the pulse that we are considering and the last part of the proof of the lack of existence of the proper causal trajectory
(which details can be found in Supplementary Material). 
The pulse takes the form
\begin{equation} \label{pulse}
{\bf E}(\x,t) = f(z-ct) {\bf e}_x
\ \ \ \ 
 {\bf B}(\x,t) = f(z-ct){\bf e}_y
\end{equation}
where $f$ is a one dimensional function. We take this function as constant equal to $f_0$ for
$  - \frac{3}{4} T \leq   t  \leq 0$ and zero otherwise.
The above describes one-dimensional pulse (as we are in 3D it is infinite in the 
$x-y$ plane) traveling along $z$.

The proof of the non-existence of physical  particle trajectory in fact is based on derivation
of lower and upper bound on final particle velocity $v_f$.
This derivation is a technical step and is described in Supplementary Material in details.
These bounds take the form
\begin{eqnarray} \label{nierownosc1}
\sqrt{3} E_{max} T^{3/2}  &\geq&   v_f
\\ \label{nierownosc2}
\frac{v_f}{\sqrt{1 - v_f^2}}   &\geq&   E_{mean} T - 3 E_{max}^2 T^3.
\end{eqnarray}
In the above $E_{max} = f_0$ is the maximal electric field of the pulse and $E_{mean} = \frac{1}{T} \left|\int_0^{T} \mbox{d} t \ {\bf E}  \right| $ is the 
absolute value of the mean electric field in the pulse. 
The Material about particle trajectory enters the above inequalities in the form of $v_f$.
Putting both inequalities together we obtain
\begin{equation} \label{wynik4}
\frac{\sqrt{3} E_{max} T^{3/2}}{\sqrt{ 1 -3 E_{max}^2 T^3 }  }  \geq  E_{mean} T - 3 E_{max}^2 T^3.
\end{equation}
In Supplementary Material we show that in the case of  the pulse considered above  we have
$ E_{max}/2  \leq E_{mean} \leq E_{max} =  f_0 $.
Having that we  clearly see that for $E_{max} \ll 1$, $T\ll 1$  the above inequality is violated.
Thus we found the contradiction needed in the ad absurdum proof. 
%This means that in the case of above mentioned parameters there does not exist a particle trajectory that enables to satisfy both energy and momentum conservation.

As we completed the proof we may now move to discussion of the above result and its connection to other works.

We now briefly discuss the lack of Lagrangian formulation of the theory we are considering.
In principle we  can write down, what is naively considered as Langrangian of classical electrodynamics and derive equations
of motion. 
Still when we solve this equation (for example in the case of static charge)
we find that the terms present in the Langrangian are infinite ($E^2$ term)
or not well defined ($A^\mu j_\mu$ -- the electromagnetic potential $A^\mu$ {\it in the point the particle is} is unknown). 
Thus the Langrangian cannot be computed and we cannot derive energy and momentum conservation from this undefined formulation.
As, one may say, it formally exists (we can write it down) still being precise it does not exist - $A^\mu j_\mu$ is an unknown quantity.
%One of the ways would be to define those quantities. 
%In our case as we do not provide the renormalization procedure valid at all times, we are not able to define the term $E^2$ present in the Langrangian
%of Hamiltonian formulation. 
%That is on of the reasons that we started directly from Maxwell equations defining the classical electrodynamics in this way.

Now we want to discuss shortly the work of  Dirac \cite{Dirac}. 
We followed Dirac formulation of the theory of point particles.
As a result we re-derived Eqs.~(\ref{enN}) and (\ref{momN}) which are 
are written in Dirac's paper in the form of an integral
over proper time $s$ i.e. $\int ds \, g_\mu(s) =0 $ (see  Supplementary Material for more details).
Dirac defines his theory by taking $g_\mu(s) = 0$. 
This choice leads to well known form of radiation reaction force known as Lorentz-Abraham-Dirac force (LAD force).  
However, at least in some cases \cite{Dirac,Rohlich}, the particle trajectories resulting from LAD force are non-causal.
This leads to serious difficulties. Dirac in his work notices that 
such behavior leads to speed of the electromagnetic signal being faster than the speed of light.
He writes
 "it is possible for a signal to be transmitted faster than light through the interior of the electron being the region of failure ... of some of the elementary properties of space and time".
We might consider another situation, placing a lot of electrons in one line, than sending the signal at those electrons one can (at least theoretically) get {\it any} speed of signal transmission.

Now we move to the discussion of how the findings of this paper imply to the investigation of the theory of extended charge models.
One of the research direction when trying to derive the radiation reaction force
was by considering the extended charge models.
There one tries to model the elementary charged particle by considering charge of finite volume.
In such model the total electromagnetic field is finite in any spatial point. 
It can be shown that the Langrangian is finite and we can use this formulation
-- as a consequence the energy and momentum is conserved with the Lorentz force  being the only electromagnetic force present.
We might add that in such models the parts of the charge repeal each other and one needs the non-electric
attractive forces to hold the charge together as a whole, preventing it from "explosion" due to electrostatic repulsion (see an example of such forces in \cite{Birula}). 
Still such models are appealing as they give clear physical mechanism (retarded effects) behind radiation reaction force. 
When the particle starts to accelerate, then due to retarded effects, the sum of all this forces is not zero - the self force arises naturally.  
In addition as the Lorentz force is the only electromagnetic force present, thus
the charged object shall start to move {\it only} when the external field touches it.
Thus the extended charge models are a good candidate to find the causal radiation reaction force and as a result the causal classical electrodynamics of point particle.
However the charged object is a composite system that possess an infinite number of internal degrees of oscillation.
This internal degrees of freedom may be excited as the external electromagnetic fields pass though the object.
If that happens the composite object {\it cannot} model the elementary particle -
its energy and momentum are {\it not} given only by total mass and velocity.
In such a case the extended charge model shall not give the causal classical electrodynamics of point particle.
We have just shown that this theory does not exists which implies that in the case of extended charge model theory the 
 internal excitation will appear (at least in the situation considered in this paper).

The above implies that when in the extended charge models one assumes lack of 
excitations of internal degrees of freedom, then such incorrect assumption
may give unphysical radiation reaction force. This is the case of 
famous extended charge model proposed by Lorentz and described
in many books \cite{Jakson,Lorentz,Abraham}. 
Lorentz considers the model of electron as a rigid body (which is inconsistent with relativity principle)
and does not take into account any internal excitations of this composite system (which is the incorrect assumption). 
His derivation of the radiation reaction force is therefore incorrect \cite{dopiskaFeynman}.
Still the obtained result is equivalent, in nonrelativistic limit, to the one obtained by Dirac.

Here we mention the works of Yaghjian \cite{Yag2} and Medina \cite{Medina}
who consider the extended charge models. Starting from causal radiation reaction force (in extended model)
they try to obtain such force for point particle, by taking the charge radius to a very small value.
However both author finds it impossible to obtain the point particle limit.
No causal radiation reaction force for point particle is found.

Here we might ask a question about the status of classical electrodynamics of point particles,
as we  have just shown that it, strictly speaking, does not exist.
The most natural (at least to the author) solution of this problem is that
classical electrodynamics is an approximation to the quantum electrodynamics.
Every approximation has the regime of parameters where it ``works", i.e. correctly describes
the undergoing processes. 
It the above proof we used the electromagnetic pulse that lasted much shorter that $1/\tau_0$.
It is known \cite{Landau} that for such pulses the quantum effect  dominate the radiation
reaction effects. This means that in this regime, one cannot simply use the classical approximation
to describe the considered process.

We need to say that there exist a regime where radiation reaction force can be treated as a perturbation to
the Lorentz force. In such regime one can derive approximate causal force.
The most popular and the one that seems to be the best physically motivated \cite{Spohn} is the Landau-Lifshitz
force \cite{Landau}.
The Newton equations with that force seems to be the best ``approximation" to the quantum electrodynamics
in the regime where purely quantum processes are negligible and
still one wants to describe the loss of energy and momenta of charged particles due to radiation effects.

{\bf Acknowledgements} \\
I greatly acknowledge Andrzej Veitia for reading of the manuscript and
creative comments that crucially influenced the final version of the paper.

{\bf Supplementary Material} is available for this paper.

\newpage

\begin{widetext}

\begin{center}
{\bf  SUPPLEMENTARY MATERIAL}
\end{center}

\section{General Considerations}

\subsection{Formulas present in the paper}

Below we rewrite the formulae present in the main body of the paper.
The energy and momentum conservation read
\begin{eqnarray}\label{enN_dod}
&&   \int \mbox{d} t \  \, {\bf E} \cdot {\bf v}  =  \gamma_f -1   +    E_{rad}  
\\ \nonumber
\\ \label{momN_dod}
&&   \int \mbox{d} t \  \left({\bf E} + {\bf v} \times {\bf B} \right) =
 {\bf v}_f \gamma_f  +   {\bf P}_{rad}. 
\end{eqnarray}
\\
\\
The pulse takes the form
\begin{equation} \label{pulse_dod}
{\bf E}(\x,t) = f(z-ct) {\bf e}_x
\ \ \ \ 
 {\bf B}(\x,t) = f(z-ct){\bf e}_y
\end{equation}
where $f$ is a one dimensional function. We take this function as constant equal to $f_0$ for
$  - \frac{3}{4} T \leq   t  \leq 0$ and zero otherwise.
The above describes one-dimensional pulse (as we are in 3D it is infinite in the 
$x-y$ plane) traveling along $z$.
\\
\\
The lower and upper bounds read
\begin{eqnarray} \label{nierownosc1_dod}
\sqrt{3} E_{max} T^{3/2}  &\geq&   v_f
\\ \label{nierownosc2_dod}
\frac{v_f}{\sqrt{1 - v_f^2}}   &\geq&   E_{mean} T - 3 E_{max}^2 T^3.
\end{eqnarray}
In the above $E_{max} = f_0$ is the maximal electric field of the pulse and $E_{mean} = \frac{1}{T} \left|\int_0^{T} \mbox{d} t \ {\bf E}  \right| $ is the 
absolute value of the mean electric field in the pulse. 
\\
\\
The final inequality 
\begin{equation} \label{wynik4_dod}
\frac{\sqrt{3} E_{max} T^{3/2}}{\sqrt{ 1 -3 E_{max}^2 T^3 }  }  \geq  E_{mean} T - 3 E_{max}^2 T^3.
\end{equation}

\subsection{Formulae for energy and momentum density}

In the main body of the paper we assumed certain form of energy and momentum
densities of electromagnetic fields. 
As it is known \cite{Jakson,Landau}  the assumed formulae are
the simplest choice of quantities that are conserved during the free evolution of
electromagnetic fields. 

Still we know that the Lorentz force is proved to be the correct force (in the case of point particles where only external fields are present)  in the cases when classical electrodynamics ``works".
Therefore one when constructing classical electrodynamics needs to effectively arrive at Lorentz force in the situations where the radiation reaction force can be practically omitted.

As we have discussed in the main body of the paper we somehow derived Lorentz force using the assumption about energy and momentum of the electromagnetic fields.
If we would choose another invariant of the Maxwell equations than we {\it would not} arrive at the Lorentz force in the formulas for energy and momentum conservation.

\subsection{Derivation of inequalities on $v_f$ given by Eqs.~(\ref{nierownosc1_dod}) and (\ref{nierownosc2_dod}).}

% W glownej czesci pracy nie ma nic o tym ze T jest maksymalnym czasem przez ktory elektron odczuwa pole. 
% Teraz potrzeba by to zalozyc. Wszystko wyprowadzic z tym zalozenie a na koncu udowodnic.

We notice that in Eq.~(\ref{pulse_dod}) the function $f$ is nonzero for the time $\frac{3}{4}T$.
In below calculation we assume that the particle experiences the pulse for time $T$ at most.
At the end of this part of Supplementary Information we show that this assumption is justified if condition
\begin{equation}\label{condition}
  E_{max} T^2 < \frac{2}{9} 
\end{equation} 
is satisfied. In the above $E_{max}$ is the maximal value of the electric field in the pulse experienced by the particle.
In the main body of the paper at last we restrict our considerations to the region $E_{max} \ll 1$ and $T \ll 1$. This restriction makes the above condition
to be satisfied.

As $T$ is the maximal time the pulse acts on the particle, thus 
 in the left hand side integrals in Eqs.~(\ref{enN_dod}) and (\ref{momN_dod}) we have $\int \mbox{d} t =  \int_0^T \mbox{d} t $ (as for other $t$ 
 the external fields acting on the particle vanish). 
But this is not the case in formulas for $ E_{rad}$ and $ {\bf P}_{rad}$ 
(given by Eqs.~(\ref{Erad2}) and (\ref{Prad0}))  where 
the acceleration and velocity may have nonzero value even for $t> T$ and there we have
$\int \mbox{d} t =  \int_0^\infty \mbox{d} t $.

We now derive an upper bound on $v_f$ using energy conservation. 
From Eq.~(\ref{enN_dod}) we get
\begin{equation}\label{eq22}
E_{max} v_{max} T   \geq \int_0^T \mbox{d} t \  \, {\bf E} \cdot {\bf v}   
\end{equation}
where  $v_{max} = v(t_{max}) $ is the maximal speed of the particle in the interval
$0 \leq t \leq T$ which takes its value for $t=t_{max}$.
As the end of the this part of Supplementary Information we derive an inequality
\begin{equation}\label{gora}
  E_{rad}  \geq  \frac{2}{3}\frac{{v_{max}}^2}{ T}.
\end{equation}
%We stress that in derivation of the above inequality we used the assumption that
%the particle starts to move {\it after} the electromagnetic pulse touches it.
From Eq.~(\ref{enN_dod}), (\ref{eq22}), (\ref{gora}) and the fact that $\gamma_f - 1 \geq \frac{v_f^2}{2}$  we have
\begin{equation}\label{glowne1}
 E_{max} v_{max} T \geq  \frac{v_f^2}{2} +   \frac{2}{3}\frac{{v_{max}}^2}{ T}. 
\end{equation}
From the above inequality we get
\begin{eqnarray*}
 E_{max} v_{max} T  \geq   \frac{2}{3}\frac{{v_{max}}^2}{ T}
\end{eqnarray*}
which gives us upper bound on $v_{max}$ that reads
\begin{equation}\label{wynik1}
  \frac{3}{2}  E_{max} T^2  \geq v_{max}. 
\end{equation}
To obtain the upper bound on $v_f$ we make use of inequality (\ref{glowne1}) to get
\begin{equation}\label{wyn11}
 E_{max} v_{max} T \geq    \frac{v_f^2}{2}.
\end{equation}
From inequalities (\ref{wynik1}) and (\ref{wyn11}) we obtain
\begin{eqnarray*}
 \frac{3}{2}  E_{max}^2 T^3    \geq E_{max} v_{max} T \geq    \frac{v_f^2}{2}.
\end{eqnarray*}
The above inequality gives the upper bound on $v_f$ that reads
\begin{equation}\label{wynik22}
\sqrt{3} E_{max} T^{3/2}  \geq   v_f.
\end{equation}
\\
\\

Now we derive the lower bound.
From Eq.~(\ref{momN}) we obtain
\begin{equation} \label{im0}
\left| \int_0^T \mbox{d} t \  \left({\bf E} + {\bf v} \times {\bf B} \right) \right| \geq  E_{mean} T  - v_{max} T E_{max} 
\end{equation}
where $E_{mean} = \frac{1}{T} \left|\int_0^{T} \mbox{d} t \ {\bf E}  \right| $
and we used the fact that in the electromagnetic wave $E = B$.
From  Eq.~(\ref{momN}) we have
\begin{eqnarray*}
&& v_f \gamma_f \geq \left| \int \mbox{d} t \  \left({\bf E} + {\bf v} \times {\bf B} \right) \right|-|P_{rad}| \geq 
\\
\\
&&
\geq E_{mean} T  - v_{max}T E_{max} - E_{rad} \geq 
\\
\\
&& \geq E_{mean} T  - 2v_{max} T E_{max}
\geq  E_{mean} T - 3 E_{max}^2 T^3
\end{eqnarray*}
where we  used inequalities given by Eqs.~(\ref{Prad0}), (\ref{eq22}), (\ref{wynik1})  and  (\ref{im0}). 
The above gives the lower bound on $v_f$ which reads
\begin{eqnarray*} 
v_f \gamma_f  \geq  E_{mean} T - 3 E_{max}^2 T^3.
\end{eqnarray*}
where $E_{mean} = \frac{1}{T} \left|\int_0^{T} \mbox{d} t \ {\bf E}  \right| $ is the 
absolute value of the mean electric field in the pulse.

\subsubsection{Derivation of inequality given by Eq.~ (\ref{gora})}

From  Eq.~(\ref{Erad2}) we have
\begin{equation}\label{ine1} 
 E_{rad} = \frac{2}{3}  \int \mbox{d} t \  \gamma^6 \left({{\bf a}}^2 -  \left( {\bf v} \times {\bf a} \right)^2    \right) \geq  \frac{2}{3}  \int \mbox{d} t \, {{\bf a}}^2 \geq  \frac{2}{3}  \int_0^{t_{max}} \mbox{d} t \, {{\bf a}}^2
\end{equation}
where $0 \leq  t_{max} \leq T$ as stated in the main text.
Now we make use of
\begin{equation}\label{ine2}
\frac{1}{t_{max}} \int_0^{t_{max}} \mbox{d}t \ {\bf a}^2
\equiv\langle {\bf a}^2 \rangle \geq \langle {\bf a} \rangle^2 \equiv 
\left( \frac{1}{t_{max}} \int_0^{t_{max}} \mbox{d}t \ {\bf a} \right)^2 = \frac{{\bf v}_{max}^2}{t_{max}^2}.  
\end{equation}
where ${\bf v}_{max} = {\bf v}(t_{max})$ as stated in the main body of the paper.
We note that in the above  use the assumption that the particle
starts to move {\it after} the electromagnetic pulse touches it i.e. 
$\int_0^{t_{max}} {\bf a} = {\bf v}(t_{max})$.
From Eqs.~(\ref{ine1}) and (\ref{ine2}) we have
\begin{equation} \label{ine3}
  E_{rad} \geq \frac{2}{3} \frac{v_{max}^2}{t_{max}} \geq  \frac{2}{3} \frac{v_{max}^2}{T}
\end{equation}
as $ 0 < t_{max} \leq T  $.

\subsubsection{Derivation of condition given by Eq.~(\ref{condition})}

We now concentrate our attention on the following problem.
Above we assumed that the pulse acts on the particle
in the interval  $ 0 \leq t \leq T$. Now we need to connect it to the shape of the pulse given
by Eq.~(\ref{pulse_dod}) where
$f$ function equal to $f_0$ for $ - \frac{3}{4}T \leq   t  \leq 0$ and zero otherwise.
We take $\x =0$ as the position of the particle for $t<0$.
Thus for $t < 0$ the particle is at rest since the pulse arrives at time $t=0$ as one can see 
from the form of $f$ function and Eq.~(\ref{pulse_dod}). 
The latest time the field can influence the particle is equal to
$T_l = \frac{3}{4} T + v_{max} \frac{3}{4} T $ as $v_{max}$ is the maximal speed
 of particle in the interval $0 \leq t \leq T$.
The demand that $T_l < T$ reads
$ \frac{3}{4} T ( 1+ v_{max} ) <  T$.  From Eq.~(\ref{wynik1}) we obtain
\begin{eqnarray*}
\frac{3}{4} T (1+v_{max}) \leq   \frac{3}{4} T  \left(  1  +  \frac{3}{2}  E_{max} T^2 \right) < T.
\end{eqnarray*}
To satisfy the above we simply need to take
\begin{eqnarray*}
   E_{max} T^2 < \frac{2}{9}. 
\end{eqnarray*}

\subsection{Calculation of $E_{mean}$}

Now we calculate  $E_{mean} = \frac{1}{T} \left|\int_0^{T} \mbox{d} t \ {\bf E}  \right| $.
We find that the time the particle experiences the pulse $T_e$ is bounded by 
$ \frac{3}{4} T(1-v_{max}) \leq T_e \leq T   $. 
 From Eq.~(\ref{wynik1}) we obtain
\begin{eqnarray*}
 \frac{3}{4} T  \left(  1 - \frac{3}{2}  E_{max} T^2 \right)   \leq \frac{3}{4} T(1-v_{max})   \leq T_e \leq T
\end{eqnarray*}
Using Eq.~(\ref{condition}) we get
\begin{eqnarray*}
\frac{1}{2} T  \leq T_e \leq T.
\end{eqnarray*}
From the definition of the pulse we find that $E_{mean} = \frac{ f_0}{T} T_e$
which together with the above gives
\begin{eqnarray*}
\frac{1}{2}  f_0   \leq E_{mean}  \leq   f_0 = E_{max}.
\end{eqnarray*}

\subsection{Connection with the work of Dirac}

In his paper \cite{Dirac} Dirac calculates the flow of electromagnetic four-momentum through the sphere of infinitesimally small radius $\epsilon$
around the charged particle.
Using the standard energy momentum tensor he arrives at expression (the units are defined in \cite{Dirac}):
\begin{equation}\label{D1}
\delta P_\mu = \int_{s_i}^{s_f} \left(  \frac{e^2}{2\epsilon}  \dot v_\mu - e v_\nu f^\nu_\mu \right) ds 
\end{equation}
where
\begin{equation}\label{D2}
f^\nu_\mu = F^\nu_{\mu,in} + \frac{2}{3} e \left(  \ddot v_\mu v^\nu - \ddot v^\nu v_\mu \right)  
\end{equation}
and $F^\nu_{\mu,in}$ denotes the electromagnetic field tensor of the incoming field (in our case this is the external electromagnetic pulse). 
In the above $s$ is the proper time. 
The quantity $\delta P_\mu$ is the change from time $s_i$ to $s_f$, of the electromagnetic four-momentum in the whole space apart from ball 
of radius $\epsilon$ around the charge. 
Now the initial and final value of the four-momentum $P_{\mu,i}$ and $P_{\mu,f}$ of the system reads
\begin{equation}\label{Pmufi}
P_{\mu,i} = P_{\mu,i,r} + P_{\mu,i,\epsilon}
\ \ \ \ \ 
P_{\mu,f} = P_{\mu,f,r} + P_{\mu,f,\epsilon}
\end{equation}
In the above $P_{\mu,i,r}$ and $P_{\mu,f,r}$ are initial and  final  four-momentum of the electromagnetic fields outside the ball of radius $\epsilon$
where as $P_{\mu,i,\epsilon}$ and $P_{\mu,f,\epsilon}$ are the initial and final four momentum of the electromagnetic plus non-electromagnetic (Poincare stresses)
fields inside the ball of radius $\epsilon$.
As mentioned above $\delta P_\mu$  is the flow of four-momentum outside the ball of radius $\epsilon$ from the initial to final situation.
Thus the final four-momentum of the part of the system outside the ball of radius $\epsilon$, $P_{\mu,f,r}$, is equal to the initial
one   $P_{\mu,i,r}$ plus $\delta P_\mu$ i.e.
\begin{equation}\label{deltaPmuW}
 P_{\mu,f,r} = P_{\mu,i,r} + \delta P_\mu.
\end{equation}
The renormalization procedure defined in this work gives
\begin{equation}\label{mmu}
m v_{\mu,i} = \frac{e^2}{2 \epsilon}   v_{\mu,i} + P_{\mu,i,\epsilon}
\ \ \ \ \ \ 
m v_{\mu,f} = \frac{e^2}{2 \epsilon}   v_{\mu,f} + P_{\mu,f,\epsilon}
\end{equation}
Starting from Eq.~(\ref{Pmufi}) we calculate
\begin{eqnarray*}
&& P_{\mu,f}- P_{\mu,i} = P_{\mu,f,r} - P_{\mu,i,r} + P_{\mu,f,\epsilon} - P_{\mu,i,\epsilon}
= P_{\mu,f,r} - P_{\mu,i,r}  +\left( m v_{\mu,f} - \frac{e^2}{2\epsilon}  v_{\mu,f} \right)
- \left( m v_{\mu,i} - \frac{e^2}{2\epsilon}  v_{\mu,i} \right)
\\
&&
= \delta P_\mu  + \left( m v_{\mu,f} - \frac{e^2}{2\epsilon}  v_{\mu,f} \right)
- \left( m v_{\mu,i} - \frac{e^2}{2\epsilon}  v_{\mu,i} \right)
\end{eqnarray*}
where we first used Eq.~(\ref{mmu}) and then Eq.~(\ref{deltaPmuW}). 
As we assume four-momentum conservation $P_{\mu,f} = P_{\mu,i} $  the above gives
\begin{equation}\label{RUCH}
0 = \delta P_\mu  + \left( m v_{\mu,f}  - \frac{e^2}{2\epsilon}  v_{\mu,f} \right)
- \left( m v_{\mu,i} - \frac{e^2}{2\epsilon}  v_{\mu,i} \right)
\end{equation}
Dirac in his work obtain exactly the same equation which in his notation reads
\begin{equation} \label{rowDD}
B_\mu(s_f) - B_\mu(s_{i})  =  \delta P_\mu =  \int_{s_{i}}^{s_f} ds  \left(  \frac{e^2}{2\epsilon} \dot v_\mu - e v_\nu f^\nu_\mu \right) 
\end{equation}
where
\begin{eqnarray*}
B_\mu = \left( \frac{ e^2}{2\epsilon}   -m \right) v_\mu.
\end{eqnarray*}
We rewrite Eq.~(\ref{rowDD}) as 
\begin{eqnarray*}
 \int_{s_{i}}^{s_f} ds \, g_\mu(s) = 0 
\ \ \ \ \
g_\mu(s) =  \frac{1}{2} e^2 \epsilon^{-1} \dot v_\mu - e v_\nu f^\nu_\mu - \dot B_\mu.
\end{eqnarray*}

To derive the equation of motion Dirac chooses the simplest way, that is he takes
\begin{eqnarray*}
 0 = g_\mu =    \frac{1}{2} e^2 \epsilon^{-1} \dot v_\mu - e v_\nu f^\nu_\mu - \dot B_\mu.
\end{eqnarray*}
By performing calculation one obtains from above the relativistic form of the LAD force.
\\
\\

On the other hand by performing the integrals in Eq.~(\ref{RUCH}) and using Eqs.~(\ref{D1}), (\ref{D2}) we obtain
\begin{eqnarray*}
&& m \left( v_\mu(s_f) - v_\mu(s_i) \right)  
=   \int_{s_{i}}^{s_f} d s \,    e  v_\nu f^\nu_\mu  
=  \int_{s_{i}}^{s_f}  d s \,    e  v_\nu  \left(   F^\nu_{\mu,in} + \frac{2}{3} e \left(  \ddot v_\mu v^\nu - \ddot v^\nu v_\mu \right)  \right)
\\
\\
&& 
= w(s_f) - w(s_{i})
+ \int_{s_{i}}^{s_f} d s \,    e  v_\nu   F^\nu_{\mu,in}
+ \frac{2}{3} e^2 \int d s \,  \dot v_\nu  \dot v^\nu v_\mu   
 \end{eqnarray*}
where $ w = \frac{2}{3} e^2  v_\nu \left(   \dot v_\mu v^\nu - \dot v^\nu v_\mu    \right) $.
As $ w(s_f) = w(s_{i}) = 0$ the above are the same equations as given by Eq.~(\ref{enN_dod}) and (\ref{momN_dod}).

In principle in order to derive Eqs.~(\ref{enN_dod}) and (\ref{momN_dod}) we could use 
Dirac's result. Still we wanted to re-calculate it in a different way. 
Instead  of calculating flow of four-momentum outside the ball of radius $\epsilon$ from the initial to final situation, we calculated the initial and final energy of the system
by integrating the four-momentum density over the entire space.
As these calculation differ significantly from the one performed by Dirac we decided
to present them in further part of Supplementary Information.

\subsection{External pulse description}

In this work we external pulse of the form (here we use SI units)
\begin{equation} \label{pulseNN}
{\bf E}_{ex}(\x,t) =c f(z-ct) {\bf e}_x
\ \ \ \ 
 {\bf B}_{ex}(\x,t) = f(z-ct){\bf e}_y.
\end{equation}
In the above we clearly notice that the pulse describes the pulse uniform in $x,y$ plane -- it is one dimensional.
Uniform Maxwell equations read
\begin{eqnarray*}
\left(  \frac{1}{c^2} \frac{\partial^2 }{\partial t^2}  - \triangle \right) {\bf E}_{ex}(\x,t) = 0
\end{eqnarray*}
and the same for ${\bf B}_{ex}$.
%In the one dimensionall case it reads
%\begin{eqnarray*}
%\left(  \frac{1}{c^2} \frac{\partial^2 }{\partial t^2}  - \frac{\partial^2}{\partial z^2} \right) {\bf E}_{ex}(z,t) = 0
%\end{eqnarray*}
Note that shape given by Eq.~(\ref{pulseNN}) is a solution of the above equation.

We might additionally add that the above pulse is special due to the fact that 
$\int dt  \, {\bf E}(\x,t) \neq 0$ -- the electric field in any point of space integrated over time in non-zero.
the standard electromagnetic pulses -- for example laser pulses are of the form $\int dt  \, {\bf E}(\x,t) = 0$.

Such kind of pulse may be generated in the following way.
We take positive charge and put it in the position $ (0,0,-R)$ (the initial position of the static point charge is $(0,0,0)$).
Then we move it with constant acceleration for time $\frac{3}{4} T$ in the $x$ direction and then set the acceleration to zero.
As a result locally around the point charge the radiated pulse will have the form given by Eq.~(\ref{pulseNN}).
In fact, in all the consideration in this work we could use just described pulse.
We would obtain exactly the same results -- simply the technical part of calculation would be a but more difficult and longer.

\section{Derivation of conservation laws given by Eqs.~(\ref{enN_dod}) and (\ref{momN_dod})}

In what follows we derive formulae for initial and final energy/momentum of the system. 
We perform this calculation as they significantly differ from the one presented by Dirac in \cite{Dirac}.

\subsection{Formal solution of electromagnetic fields}

Before the derivation of the energy and momentum of the electromagnetic fields we 
remind the situation that we analyze.
We assumed that initially the charged point particle is at rest.
Then an electromagnetic pulse reaches it  and acts on the particle for finite time, eventually
leaving it. Below we use SI units.

The formal solution of this problem reads 
\begin{equation}
{\bf  E}_{tot} (\x,t) = {\bf E}_{ex}(\x,t) + {\bf E}_{ch}(\x,t)
\end{equation}
and the same for ${\bf B}_{tot}$. The above ${\bf  E}_{tot}$, ${\bf E}_{ex}$,  ${\bf E}_{ch}$ denotes the total, external pulse and generated by the charge
electric field respectively. 
The field of the pulse are given by Eq.~(\ref{pulseNN}) whereas the field generated by the charge is given by \cite{Jakson}
\begin{equation} \label{Ech}
{\bf E}_{ch} = -\nabla \Phi - \partial_t {\bf A}
\ \ \ \
{\bf B}_{ch} = \nabla \times {\bf A}
\end{equation}
where
\begin{eqnarray} \label{Phi}
&& \Phi(\x,t) = \frac{1}{4\pi \epsilon_0} \int d \x' dt' \, G(\x-\x',t-t') \rho(\x',t')
\\ \nonumber
\\ \label{A}
&&  {\bf A}(\x,t) =  \frac{\mu_0}{4\pi} \int d \x' dt' \, G(\x-\x',t-t') {\bf J}(\x',t')
\\ \nonumber
\\ \label{Gdef}
&& G(\x,t) = \frac{\delta(t- |\x|/c)}{|\x|}.
\end{eqnarray}
Note above the presence of retarded Greens function and lack of the advanced one.
As it is known any linear combination of advanced and retarded Greens function is the solution of 
Maxwell equations \cite{Jakson}. Still this combination is uniquely determined by the initial conditions
- the values of electric and magnetic fields at initial time.
In our case the initial conditions (sourceless pulse  and static coulomb field) are such that no advanced Green function is possible. Initial conditions are satisfied when retarded Greens function is used.
\\

\subsection{Initial and final energy and momentum - formulas}

Having specify the fields we now move to energy and momentum calculation.
%Now we derive the mathematical expressions representing energy and momentum conservation.
The energy and momenta of electromagnetic fields  are quadratic function of the fields which we denote as
${\cal E}(F,F)  $ and ${\bf P}(F,F)$ where $F$ denotes the field in general (${\bf E}$, ${\bf B}$). 
Both of these functions are linear in both variables.
Long after the pulse left the particle the electromagnetic fields  are  composed  of three  parts: 
\begin{itemize}
\item fields of the pulse $F_{a,f}$
\item fields attached to the charged particle $F_{b,f}$
\item fields emitted (radiated) by the accelerated particle $F_{c,f}$.
\end{itemize}
The total ``final" field is equal to $F_f = F_{a,f}+F_{b,f}+F_{c,f}$
with the energy $ {\cal E}(F_f,F_f)$.
Now we use the renormalization procedure subtracting from above
the energy of the fields attached to the particle ${\cal E}( F_{b,f},F_{b,f})$
and adding the total particle energy equal to $\gamma_f mc^2$
where $\gamma_f = (1-v_f^2/c^2)^{-1/2}$  and ${\bf v}_f$ is the final velocity of the particle.
As a result the final energy  ${\cal E}_f$ of the system is equal to
\begin{equation}\label{Efinal}
{\cal E}_f =  {\cal E}(F_f, F_f) - {\cal E}( F_{b,f},F_{b,f}) + \gamma_f mc^2.
\end{equation}
Using the same reasoning we obtain that final momentum of the system ${\bf P}_f$ reads
\begin{equation}\label{Pfinal}
{\bf P}_f =  {\bf P}(F_f, F_f) - {\bf P}( F_{b,f},F_{b,f}) + \gamma_f m {\bf v}_f.
\end{equation}
Now we turn our attention to the initial state - long before the pulse touches the particle.
In such a case we deal only with $F_{a,i}$ and $F_{b,i}$. The fields radiated by the charge particle 
$F_{c,i}$ do not exists. Therefore we have $F_i = F_{a,i} + F_{b,i} $.
Using again the renormalization procedure we obtain that the initial energy ${\cal E}_i$ and momentum ${\bf P}_i$ of the system read
\begin{eqnarray}\label{Einitial}
&& {\cal E}_i = {\cal E}(F_i, F_i) - {\cal E}( F_{b,i},F_{b,i}) +  mc^2
\\ \label{Pinitial}
&& {\bf P}_i = {\bf P} (F_i, F_i) - {\bf P}( F_{b,i},F_{b,i}). 
\end{eqnarray}
%In Supplementary Information we perform analytical calculation of initial and final energies ${\cal E}_i$, 
%${\cal E}_f$  and momenta ${\bf P}_i$, ${\bf P}_f$.
%By equating  ${\cal E}_i = {\cal E}_f$  and ${\bf P}_i = {\bf P}_f$ we obtain 
%(calculations performed in Supplementary Information)

%We are interested in the energy and momentum of the electromagnetic fields
%long before and long after the pulse acts on the particle at times $t_i$ and $t_f$ respectively.

\subsection{Renormalization procedure discussion}

We now discuss the use of renormalization procedure performed above.
We clearly see that we subtracted the energy of the attached fields generated by the particle.
It was performed in the initial and final situation.
In such a case the point particle moves with constant velocity (or is static) for very long time.

Here we discuss what ``very long time" means.
If the charge is at rest for time $T_s$ than
the static coulomb field is present in space up to radius $c T_s$.
For $r > r_s$ the field may be different than coulomb field (there might exist field radiated by the particle).
In the renormalization procedure we subtract the energy (here we deal with static charge thus the momentum is zero)
of the coulomb electric field. 
The energy of the coulomb field in the space apart from ball of radius $cT_s$ is equal to $mc^2 \frac{r_0}{cT_s}$.
This is the maximal ``error" of the renormalization procedure. As in the calculation, at the end, we take the limit $T_s \rightarrow \infty$
the error disappears.

\subsection{Results of the calculation}

In the next two sections we calculate the energy and momentum of the system. Here we briefly discuss the results
of these calculations.
The energy initial and final energy and momentum reads
\begin{eqnarray*} \nonumber
&& {\cal E}_i =   {\cal E}_{ex} +  {\cal E}_{in}(t_i)  +  mc^2
\ \ \ \ \ \ 
 {\cal E}_f =     {\cal E}_{ex}   +  {\cal E}_{in}(t_f)   + \widetilde E_{rad}  + \gamma_f m c^2
\\
\\
&& 
{\bf P}_i =   {\bf P}_{ex} +  {\bf P}_{in}(t_i)  
\ \ \ \ \ \ 
 {\bf P}_f =     {\bf P}_{ex}   +  {\bf P}_{in}(t_f)   + {\bf P}_{rad}  + \gamma_f m {\bf v}_f  
\end{eqnarray*}
where $E_{rad} = {\cal E}(F_{c,f},F_{c,f})$, ${\bf P}_{rad} ={\bf P}(F_{c,f},F_{c,f}) $ is the energy and momentum 
 of the field radiated by the particle and
$ {\cal E}_{ex} = {\cal E}(F_{a,f},F_{a,f}) =    {\cal E}(F_{a,i},F_{a,i})  $, 
$ {\bf P}_{ex} = {\bf P}(F_{a,f},F_{a,f}) =    {\bf P}(F_{a,i},F_{a,i})  $ is the energy and momentum of the 
external pulse (which is constant in time).
In addition we notice 
\begin{eqnarray*}
&& {\cal E}_{in}(t) = \epsilon_0  \int d \x \, \left( {\bf E}_{ex}(\x,t) {\bf E}_{ch}(\x,t) + c^2 {\bf B}_{ex}(\x,t) {\bf B}_{ch}(\x,t) \right)
\\
&& {\bf P}_{in}(t) =  \epsilon_0 \int d \x \, \left( {\bf E}_{ex}(\x,t) \times {\bf B}_{ch}(\x,t)
+ {\bf E}_{ch}(\x,t) \times {\bf B}_{ex}(\x,t) \right)
\end{eqnarray*}
which denote the energy and momentum of the electromagnetic fields resulting from interference of the field generated by the particle
with the field of the external pulse. Those quantities are calculated at initial and final time $t_i$ and $t_f$ and they read
\begin{eqnarray*}
&&  {\cal E}_{in}(t_i)  = 0 
\ \ \ \ \ \
 {\cal E}_{in}(t_f)  =  -  q \int_{-\infty}^\infty  dt' \,   {\bf v}(t') \cdot  {\bf E}_{ex}(\x(t'),t')
\\
\\
&&  {\bf P}_{in}(t_i) = \frac{3}{8} q T  f_0   {\bf e}_x
\ \ \ \ \ \
{\bf P}_{in}(t_f) = \frac{3}{8} q T  f_0   {\bf e}_x - q \int_{-\infty}^{\infty}  dt  \,   \left( {\bf E}_{ex}(\x(t),t) + {\bf v}(t) \times {\bf B}_{ex}(\x(t),t) \right) 
\end{eqnarray*}
In the above we notice the term $ {\bf P}_{in}(t_i) = \frac{3}{8} q T  f_0   {\bf e}_x $ which is a nonzero momentum coming from the interference between the initial 
coulomb field of the charge with the external pulse.
As a result we obtain
\begin{eqnarray*}\nonumber
&& {\cal E}_i =   {\cal E}_{ex} +   mc^2 
\\  
\\ \nonumber
&& {\cal E}_f =     {\cal E}_{ex}   + \widetilde E_{rad}  + \gamma_f mc^2   -  q \int_{-\infty}^\infty  dt' \,   {\bf v}(t') \cdot  {\bf E}_{ex}(\x(t'),t')
\\
\\ \nonumber 
&& {\bf P}_i =   {\bf P}_{ex} +  \frac{3}{8} q T  f_0   {\bf e}_x
\\ 
\\ \nonumber
&& {\bf P}_f =     {\bf P}_{ex}     + {\bf P}_{rad}  + \gamma_f m {\bf v}_f   +  \frac{3}{8} q T  f_0   {\bf e}_x    - q \int_{-\infty}^{\infty}  dt  \,   \left( {\bf E}_{ex}(\x(t),t) + {\bf v}(t) \times {\bf B}_{ex}(\x(t),t) \right) 
\end{eqnarray*}
By equating initial and final energy/momentum we obtain
\begin{eqnarray}\label{En0}
&& q \int_{-\infty}^\infty  dt' \,   {\bf v}(t') \cdot  {\bf E}_{ex}(\x(t'),t')   =    E_{rad}  + (\gamma_f  -1) mc^2    
\\ \nonumber
\\ \label{P0}
&& 
q \int_{-\infty}^{\infty}  dt  \,   \left( {\bf E}_{ex}(\x(t),t) + {\bf v}(t) \times {\bf B}_{ex}(\x(t),t) \right)  =
 {\bf P}_{rad}  + \gamma_f m {\bf v}_f  
\end{eqnarray}
For the convenience of the calculation we now rewrite the conservation laws given by above equations 
in the units  $\tau = \frac{r_0}{c}$ where  $ r_0 = \frac{\mu_0 q^2}{4 \pi m}$ is
the quantity described previously.
We now have   ${\bf r} = {\bf r}/r_0$ , $t= t/\tau$ (consequently ${\bf v}  = {\bf v}/c$), 
$ {\bf E} = {\bf E}_{ex} \frac{r_0 q}{mc^2} $, ${\bf B} = {\bf B}_{ex} \frac{r_0 q}{mc} $. 
In new units, energy and momentum conversation given by Eqs.~(\ref{En0}) and (\ref{P0})  take the form
\begin{eqnarray*}
&&   \int \mbox{d} t \  \, {\bf E} \cdot {\bf v}  =  \gamma_f -1   +    E_{rad}  
\\ \nonumber
\\ 
&&   \int \mbox{d} t \  \left({\bf E} + {\bf v} \times {\bf B} \right) =
 {\bf v}_f \gamma_f  +  {\bf P}_{rad} 
\end{eqnarray*}
where $E_{rad} = \widetilde E_{rad}/mc^2$ and $ {\bf P}_{rad} = \widetilde {\bf P}_{rad}/mc$.
The formula for the radiated energy is well known and in the new units reads \cite{Jakson}
\begin{equation} \label{Erad2}
 E_{rad} = \frac{2}{3}  \int \mbox{d} t \  \gamma^6 \left({\bf a}^2 -  \left( {\bf v} \times {\bf a} \right)^2    \right).
\end{equation}
Additionally one can derive the inequality 
\begin{equation}\label{Prad0}
 E_{rad} \geq | {\bf P}_{rad}|.
\end{equation}

\section{Calculation of the energy of the system}

\subsection{Initial and final energy - further analysis}

In what follows we denote $t_i$ and $t_f$ as ``initial" and ``final" time  when we calculate the energy and momentum of the system.
Substituting $F_f = F_{a,f} + F_{b,f} + F_{c,f}$ and $F_i = F_{a,i} + F_{b,i}$  into 
Eqs.~(\ref{Einitial}) and (\ref{Efinal}) we obtain
\begin{eqnarray*}
&& {\cal E}_i =   {\cal E}(F_{a,i},F_{a,i}) +  {\cal E}(F_{a,i},F_{b,i}) + {\cal E}(F_{b,i},F_{a,i}) +  mc^2
\\
&& {\cal E}_f =     {\cal E}(F_{a,f},F_{a,f})
+  {\cal E}(F_{a,f},F_{b,f}+F_{c,f}) + {\cal E}(F_{b,f}+F_{c,f},F_{a,f}) 
+ {\cal E}(F_{b,f},F_{c,f}) +  {\cal E}(F_{c,f},F_{b,f}) +  {\cal E}(F_{c,f},F_{c,f}) + \gamma_f m c^2
\end{eqnarray*}
We notice that the energy of the pulse does not change i.e. 
$  {\cal E}(F_{a,f},F_{a,f}) =    {\cal E}(F_{a,i},F_{a,i}) = {\cal E}_{ex}$.
Additionally  $F_{b,f}$ scales like $1/r^2$ and $F_{c,f}$ like $1/r$
and looking at the pulse we notice that $F_{c,f}$ has finite width in space. Therefore 
$ \left({\cal E}(F_{b,f},F_{c,f}) + {\cal E}(F_{c,f},F_{b,f})\right) $ goes to zero
as $r$ goes to infinity. As a result we obtain
\begin{eqnarray*}
&& {\cal E}_i =   {\cal E}_{ex}+  {\cal E}(F_{a,i},F_{b,i}) + {\cal E}(F_{b,i},F_{a,i}) +  mc^2
\\
&& {\cal E}_f =     {\cal E}_{ex}
+  {\cal E}(F_{a,f},F_{b,f}+F_{c,f}) + {\cal E}(F_{b,f}+F_{c,f},F_{a,f})  +  {\cal E}(F_{c,f},F_{c,f}) + \gamma_f m c^2 
\end{eqnarray*}
We rewrite the above as
\begin{eqnarray} \nonumber
&& {\cal E}_i =   {\cal E}_{ex} +  {\cal E}_{in}(t_i)  +  mc^2
\\ \label{EnergyWWW}
\\ \nonumber
&& {\cal E}_f =     {\cal E}_{ex}   +  {\cal E}_{in}(t_i)   + \widetilde E_{rad}  + \gamma_f m c^2
\end{eqnarray}
where $\widetilde E_{rad} = {\cal E}(F_{c,f},F_{c,f})$ is the energy of the field radiated by the particle and
\begin{eqnarray*} 
&&    {\cal E}_{in}(t_f)  =  {\cal E}(F_{a,f},F_{b,f}+F_{c,f}) + {\cal E}(F_{b,f}+F_{c,f},F_{a,f}) 
\\
&& 
 {\cal E}_{in}(t_i) = {\cal E}(F_{a,i},F_{b,i}) + {\cal E}(F_{b,i},F_{a,i}).
\end{eqnarray*}
are the energy terms  coming from the interference of the 
field of the pulse $F_{a,i}$ or $F_{a,f}$ with the field generated by the particle $F_{b,i}$ or $F_{b,f}+ F_{c,f}$.
Here it is {\it crucial} that in those terms we find {\it total} field generated by the particle - they  {\it do not} distinguish the attached and detached part.
Due to this fact, and using the formula for the energy density of the electromagnetic field, we write them as
\begin{equation}\label{Ein}
  {\cal E}_{in}(t) = \epsilon_0  \int d \x \,
\left( {\bf E}_{ex}(\x,t) {\bf E}_{ch}(\x,t) + c^2 {\bf B}_{ex}(\x,t) {\bf B}_{ch}(\x,t) \right)
\end{equation}
where we put $t=t_i$ or $t=t_f$.

\subsection{Calculation of  ${\cal E}_{in,1}$ }

We divide ${\cal E}_{in} = {\cal E}_{in,1} + {\cal E}_{in,2}$  where
\begin{equation}\label{EE}
 {\cal E}_{in,1}(t) = \epsilon_0 \int d \x \, {\bf E}_{ex}(\x,t) {\bf E}_{ch}(\x,t)
= \epsilon_0  \int d \x \,  {\bf E}_{ex}(\x,t) \left( -\nabla \Phi (\x,t) - \partial_t {\bf A}(\x,t) \right)
\end{equation}
which we calculate for $t=t_i$ and $t=t_f$. In the above we used Eq.~(\ref{Ech}).
We again divide ${\cal E}_{in,1} = {\cal E}_{in,1,1} +  {\cal E}_{in,1,2}$
where 
\begin{equation}\label{warunek1}
{\cal E}_{in,1,1}(t) = - \epsilon_0  \int d \x \,   {\bf E}_{ex}(\x,t) \cdot   \nabla_\x  \Phi(\x,t)
=   - \epsilon_0  \int d \x \,  \left( \nabla_\x \cdot \left(  {\bf E}_{ex}(\x,t)  \Phi(\x,t) \right)
-   \Phi(\x,t)  \nabla_\x \cdot {\bf E}_{ex}(\x,t)  \right).
\end{equation}
The pulse that we consider in the main body of the paper (given by Eq.~(\ref{pulseNN})) is sourceless, i.e. $\nabla_\x \cdot {\bf E}_{ex} = 0 $.
Therefore the second part of the right-hand side of the  above expression is zero.
The first part gives the surface integral. 
We take the boundary as a cylinder with initial position of the static point particle in its center and its height pointing in the $z$ direction.
The radius of this cylinder and its  height   for $t=t_i,t_f$ is much than $ct_i,ct_f$.
In such a case the electrostatic potential on the boundary of the cylinder is given by its initial value being the potential of the initially static particle
and equal to   $\Phi(\x,t) = \frac{q}{4\pi \epsilon_0  r}$. In such a case, as it can be seen, due to symmetry we have
 $\int d \x \,  \nabla_\x \cdot \left(  {\bf E}_{ex}(\x,t)  \Phi(\x,t) \right) = \int d S \,  {\bf E}_{ex}(\x,t)  \Phi(\x,t)  $ where $S$ denotes the boundary of the cylinder.
Thus 
\begin{equation}\label{Ein11}
{\cal E}_{in,1,1}(t_i) = 0 \ \ \ \ \ {\cal E}_{in,1,1}(t_f) = 0.
\end{equation}
From Eq.~(\ref{EE})
\begin{eqnarray} \nonumber
&& 
{\cal E}_{in,1,2}(t)
=
- \epsilon_0  \int d \x \,  {\bf E}_{ex}(\x,t)  \partial_t {\bf A}(\x,t) 
= 
 - \frac{\epsilon_0\mu_0}{4\pi}  \int d \x \,  {\bf E}_{ex}(\x,t) 
\partial_t \int d \x' dt' \,  {\bf J}(\x',t')  G(\x-\x',t-t')
\\ \label{row22}
&&
= 
 - \frac{1}{4\pi c^2}  
\int d \x' dt' \,  {\bf J}(\x',t')   \int d \x \,  {\bf E}_{ex}(\x,t)   \partial_t    G(\x-\x',t-t')
\end{eqnarray}
where we used Eqs.~(\ref{A}).
We assumed that initially the particle is stationary. This means that the initial current is zero.
As a result the above integral is zero for $t= t_i$ i.e.
\begin{equation}\label{Ein12In}
{\cal E}_{in,1,2}(t_i) = 0
\end{equation}
Thus we are interested only in $t = t_f$ calculation.
We have
\begin{equation}\label{cz1}
{\cal E}_{in,1,2}(t_f) = - \frac{1}{4\pi c^2}  
  \int d \x' \int_{-\infty}^{t_f} dt' \,  {\bf J}(\x',t')   \int d \x \,  {\bf E}_{ex}(\x,t_f)    \partial_{t_f} G(\x-\x',t_f-t')
\end{equation}
where in the above we took $t_f$ as the upper value of the integral over $t'$ - the integral gives the same result as its upper value would be $\infty$.
One of the possibilities to calculate this integral is to use the decomposition
\begin{eqnarray} \label{decE}
&& {\bf E}_{ex}(\x,t) = \int d\K \, 
\left( e^{-i ck t + i \K\x} {\bf E}_-(\K) + e^{i ck t + i \K\x} {\bf E}_+(\K)     \right)
\\ \nonumber
\\ \label{decB}
&& {\bf B}_{ex}(\x,t) = \int d\K \, \left(e^{-i ck t + i \K\x} {\bf B}_-(\K) + e^{i ck t + i \K\x} {\bf B}_+(\K)    \right)
\end{eqnarray}
and $\K \times {\bf E}_\pm = \mp kc {\bf B}_\pm$.
Inserting the decomposition given by Eq.~(\ref{decE}) into Eq.~(\ref{cz1}) we arrive at
\begin{eqnarray} \nonumber
&&\int d \x \,  {\bf E}_{ex}(\x,t_f)  \partial_{t_f}    G(\x-\x',t_f-t')
= \int d \x \,  \int d\K \, 
\left( e^{-i ck t_f + i \K\x} {\bf E}_-(\K) + e^{i ck t_f + i \K\x} {\bf E}_+(\K)     \right)   \partial_{t_f} G(\x-\x',t_f-t')
\\ \label{WAZNE}
&&
= 
\int d\K \, 
\left( e^{-i ck t_f } {\bf E}_-(\K) + e^{i ck t_f } {\bf E}_+(\K)     \right)   \partial_{t_f} \int d \x \,  e^{i\K\x} G(\x-\x',t_f-t')
\end{eqnarray}
% Wszystkie calki moga byc w skonczonych granicach wiec jest ok
%
%
Now we calculate
\begin{eqnarray} \nonumber
&& \int d \x \,
G(\x-\x',t_f-t')  e^{i\K\x} =  e^{i\K\x'} \int d \x -\x' \,
G(\x-\x',t_f-t')  e^{i\K(\x-\x')}   = e^{i\K\x'}
\int d (\x -\x') \, e^{i\K(\x-\x')} \frac{\delta(t_f-t'- |\x-\x'|/c)}{|\x-\x'|}
\\ \label{Gf}
&&
= e^{i\K\x'}  4\pi \int r^2 dr  \frac{\sin kr}{kr} \frac{\delta(t_f-t'-r/c)}{r} 
=   e^{i\K\x'} 4\pi \int_0^\infty dr  \frac{\sin kr}{k} \delta(t_f-t'-r/c)
=  e^{i\K\x'} 4\pi \frac{c}{k} \sin \left( kc (t_f-t')  \right)
\end{eqnarray}
where we made use of Eq.~(\ref{Gdef}).
From Eqs.~(\ref{WAZNE}), (\ref{Gf}) and (\ref{decE}) we get
\begin{eqnarray}\nonumber
&& \int d \x \,  {\bf E}_{ex}(\x,t_f)    \partial_{t_f}  G(\x-\x',t_f-t')= 
\int d\K \, 
\left( e^{-i ck t_f + i \K \x'} {\bf E}_-(\K) + e^{i ck t_f + i \K \x' } {\bf E}_+(\K)     \right)   \partial_{t_f} 4\pi \frac{c}{k} \sin \left( kc (t_f-t')  \right)
\\ \label{WAZNE2}
&&
= 4\pi c^2 \int d\K \, 
\left( e^{-i ck t_f + i \K \x'} {\bf E}_-(\K) + e^{i ck t_f + i \K \x' } {\bf E}_+(\K)     \right)   \cos \left( kc (t_f-t')  \right)
= 2\pi c^2 \left( {\bf E}_{ex}(\x',2t_f-t') + {\bf E}_{ex}(\x',t')  \right) 
\end{eqnarray}
Inserting into Eq.~(\ref{cz1}) the results of Eq.~(\ref{WAZNE2}) we obtain
\begin{equation} \label{Ein12Fi}
{\cal E}_{in,1,2}(t_f) 
=   - 
\frac{1}{2}  \int d \x'  \int_{-\infty}^{t_f} dt' \,  {\bf J}(\x',t')
\left( {\bf E}_{ex}(\x',2t_f-t') + {\bf E}_{ex}(\x',t')  \right). 
\end{equation}
\\
From Eqs.~(\ref{Ein11}), (\ref{Ein12In}) and (\ref{Ein12Fi}) we obtain
\begin{equation}\label{Ein1}
{\cal E}_{in,1}(t_i) = 0 
\ \ \ \ \ \
{\cal E}_{in,1}(t_f)
=  - 
\frac{1}{2}  \int d \x'  \int_{-\infty}^{t_f} dt' \,  {\bf J}(\x',t')
\left( {\bf E}_{ex}(\x',2t_f-t') + {\bf E}_{ex}(\x',t')  \right). 
\end{equation}

\subsection{Calculation of  ${\cal E}_{in,2}$ }

Now we calculate the second part of ${\cal E}_{in}$ that is
\begin{equation}\label{Ein2}
{\cal E}_{in,2} (t)
=
\int d \x \epsilon_0 c^2 
{\bf B}_{ex}(\x,t)
{\bf B}_{ch}(\x,t)
= \frac{1}{4\pi}
\int d \x  \,
{\bf B}_{ex}(\x,t)  
 \int d \x' \int_{-\infty}^{t_f}  dt' \,   
\nabla_{\x} \times  \left( {\bf J}(\x',t') 
G(\x-\x',t-t') \right)
\end{equation}
where we used  Eqs.~(\ref{Ech}) and (\ref{A}) and again introduced $t_f$ instead of $\infty$ as the upper limit of integral of $t'$.
As before we find that as the initial current is zero the above integral vanishes for $t=t_i$, i.e.
\begin{equation}\label{Ein2In}
{\cal E}_{in,2} (t_i) = 0
\end{equation}
Thus we are only interested in $t=t_f$.We have
\begin{eqnarray*}
{\cal E}_{in,2} (t_f) &=&
\frac{1}{4\pi}
\int d \x  \,
{\bf B}_{ex}(\x,t_f)  
 \int d \x' \int_{-\infty}^{t_f}  dt' \,   
\nabla_{\x} \times  \left( {\bf J}(\x',t') 
G(\x-\x',t_f-t') \right)
\\
\\
&& = 
 \frac{1}{4\pi} \int d \x' \int_{-\infty}^{t_f} dt' 
\int d \x \,
\left( {\bf J}(\x',t') 
G(\x-\x',t_f-t') \right)  \nabla_\x \times 
{\bf B}_{ex}(\x,t_f)  
\\
\\
&& -
\frac{1}{4\pi} 
\int d \x \,
\nabla_\x  \cdot \left( \int d \x' \int_{-\infty}^{t_f} dt' \,
 G(\x-\x',t_f-t') {\bf J}(\x',t')   \times 
{\bf B}_{ex}(\x,t_f)  \right)
\\
\\
&=& \frac{1}{4\pi c^2} \int d \x' \int_{-\infty}^{t_f} dt' 
\int d \x \,
\left( {\bf J}(\x',t') 
G(\x-\x',t_f-t') \right)  \frac{\partial {\bf E}_{ex}(\x,t_f)}{\partial t_f}
\end{eqnarray*}
where we used ${\bf a} \cdot ( \nabla \times {\bf b}) = {\bf b} \cdot (\nabla \times {\bf a} ) 
- \nabla \cdot ({\bf a} \times {\bf b})$ and Maxwell equation $\nabla \times {\bf B}_{ex} = \frac{1}{c^2}\frac{\partial {\bf E}_{ex}}{\partial t}$.
We find that the boundary term equal to
\begin{eqnarray*}
\frac{1}{4\pi} 
\int d \x \,
\nabla_\x  \cdot \left( \int d \x' \int_{-\infty}^{t_f} dt' \,
 G(\x-\x',t_f-t') {\bf J}(\x',t')   \times 
{\bf B}_{ex}(\x,t_f)  \right) = 0
\end{eqnarray*}
vanishes.
This can be seen by taking volume being a box of length larger than $ 2 c t_f$ - then the term  $  \int d \x' \int_{-\infty}^{t_f} dt' \,
 G(\x-\x',t_f-t') {\bf J}(\x',t')$  shall be zero on the boundary of the box - the speed of light is not enough to reach the boundary in time $t_f$. 
Now we again use the fourier decomposition given by Eq.~(\ref{decE}) and Eq.~(\ref{Gf}) to obtain
\begin{eqnarray} \nonumber
{\cal E}_{in,2} (t_f) &=& \frac{1}{4\pi c^2} \int d \x'  \int_{-\infty}^{t_f} dt' 
\int d \x \,
 {\bf J}(\x',t') 
G(\x-\x',t_f-t') 
\int d\K \, 
\left( - ick e^{-i ck t_f + i \K\x} {\bf E}_-(\K) + ick e^{i ck t_f + i \K\x} {\bf E}_+(\K)     \right)
\\ \nonumber
\\ \nonumber
&=&
 \int d \x' \int_{-\infty}^{t_f} dt' \, 
 {\bf J}(\x',t') 
\int d\K \, 
\left( - ick e^{-i ck t_f + i \K\x'} {\bf E}_-(\K) + ick e^{i ck t_f + i \K\x'} {\bf E}_+(\K)     \right)
\frac{1}{ck} \sin (kc (t_f-t'))
\\ \nonumber
\\ \label{Ein2Fi}
&=&
 \frac{1}{2}  \int d \x' \int_{- \infty}^{t_f} dt' \,   
 {\bf J}(\x',t') \left( - {\bf E}_{ex}(\x',t') +  {\bf E}_{ex}(\x',2t_f-t') \right).
\end{eqnarray}

\subsection{Formulas for  ${\cal E}_{in}$ }

From Eqs.~(\ref{Ein1}),  (\ref{Ein2In}) and   (\ref{Ein2Fi})
we find
\begin{eqnarray*}
 {\cal E}_{in}(t_f) = -  \int d \x'  \int_{-\infty}^{t_0} dt' \,  {\bf J}(\x',t') {\bf E}_{ex}(\x',t').   
\ \ \ \  
 {\cal E}_{in}(t_i) =0.
\end{eqnarray*}
In the case of point particle ${\bf J}(\x',t') = q {\bf v}(t')  \delta(\x' - \x(t'))$ we obtain
\begin{equation}\label{Energy3}
 {\cal E}_{in}(t_f)   = -  q \int_{-\infty}^\infty  dt' \,   {\bf v}(t') \cdot  {\bf E}_{ex}(\x(t'),t')
\ \ \ \  
 {\cal E}_{in}(t_i) =0.
\end{equation}

\subsection{Formulas for initial and final momentum of the system}

As a result from Eqs.~(\ref{EnergyWWW}) and (\ref{Energy3}) we obtain
\begin{eqnarray}\nonumber
&& {\cal E}_i =   {\cal E}_{ex} +   mc^2 
\\  \label{EnergyWynik}
\\ \nonumber
&& {\cal E}_f =     {\cal E}_{ex}   + E_{rad}  + \gamma_f mc^2   -  q \int_{-\infty}^\infty  dt' \,   {\bf v}(t') \cdot  {\bf E}_{ex}(\x(t'),t')
\end{eqnarray}

\section{Calculation of the momentum of the system}

\subsection{Initial and final momentum - further analysis}

Repeating the same steps as we did in the case of energy we obtain
\begin{eqnarray}\label{PinitialNN}
&& {\bf P}_i =   {\bf P}_{ex} +  {\bf P}_{in}(t_i)  
\\ \nonumber
\\ \label{PfinalNN}
&& {\bf P}_f =     {\bf P}_{ex}   +  {\bf P}_{in}(t_f)   + \widetilde {\bf P}_{rad}  + \gamma_f m {\bf v}_f  
\end{eqnarray}
where ${\bf P}_{ex}$ is the momentum of the external pulse, $\widetilde {\bf P}_{rad}$  the momentum of the fields radiated by the point charge and  
\begin{equation}\label{PinNN}
{\bf P}_{in}(t) =  \epsilon_0 \int d \x \, \left( {\bf E}_{ex}(\x,t) \times {\bf B}_{ch}(\x,t)
+ {\bf E}_{ch}(\x,t) \times {\bf B}_{ex}(\x,t) \right)
\end{equation}
where $t=t_i$ or $t=t_f$.

\subsection{Preliminary calculation of $ {\bf P}_{in}$}

We have
\begin{eqnarray} \nonumber
 {\bf P}_{in}(t) &=& \epsilon_0 \int d \x \, \left( {\bf E}_{ex} \times {\bf B}_{ch}
+ {\bf E}_{ch} \times {\bf B}_{ex} \right)
= 
 \epsilon_0 \int d \x \, \left( {\bf E}_{ex} \times ( \nabla \times {\bf A} ) +
(-\partial_t {\bf A}  - \nabla \Phi ) \times {\bf B}_{ex}  \right)
\\ \nonumber
\\  \label{EB}
&=&
-
 \epsilon_0 \int d \x \,  \left(  {\bf A} \times ( \nabla \times {\bf E}_{ex} )
-  {\bf E}_{ex}(\nabla \cdot {\bf A})
+ (\partial_t {\bf A}  + \nabla \Phi ) \times {\bf B}_{ex}  \right)
\end{eqnarray}
where we used Eq.~(\ref{Ech}) and additionally
\begin{equation} \label{chwila1}
 {\bf E}_{ex} \times ( \nabla \times {\bf A} )
+ {\bf A} \times ( \nabla \times {\bf E}_{ex} )
= \nabla ( {\bf E}_{ex} \cdot {\bf A} )
- \sum_i \partial_i (E_{ex,i} {\bf A})
+ {\bf A} (\nabla \cdot {\bf E}_{ex})
- \sum_i \partial_i (  A_i {\bf E}_{ex})
+ {\bf E}_{ex}(\nabla \cdot {\bf A})
\end{equation}
and found that boudary terms
\begin{eqnarray*}
\int  d \x \,  \nabla ( {\bf E}_{ex} \cdot {\bf A})    -  \sum_{i=1}^3 \partial_i ( E_{ex,i} {\bf A} +   A_i {\bf E}_{ex} )    = 0
\end{eqnarray*}
are equal to zero. This can be noticed by taking the volume of integration as a cube of box length larger than $2c|t|$ so 
that ${\bf A} =0$ on the boundary of the cube (during the time $ct$ the information that there was nonzero current shall not reach the 
volume's boundary).
Additionally in the Eq.~(\ref{chwila1}) we used the fact that our impulse is sourceless, i.e. $\nabla \cdot {\bf E}_{ex} =0 $.
% To ze jest rowne zero to jest dosc jasne zawsze bo w A jest prad i idzie z predkosci swiatla.
We divide  
\begin{equation}\label{Pindivision}
{\bf P}_{in} = {\bf P}_{in,1} + {\bf P}_{in,2} 
\end{equation}
given by Eq.~(\ref{EB}) 
where 
\begin{eqnarray}\label{Pin1}
 {\bf P}_{in,1}(t) =
- \epsilon_0 \int d \x \, \left( 
{\bf A} \times ( \nabla \times {\bf E}_{ex} ) 
+ \partial_t {\bf A}  \times {\bf B}_{ex} \right)
=  -
 \epsilon_0 \int d \x \, \left( 
- {\bf A} \times   \partial_t{\bf B}_{ex} 
+ \partial_t {\bf A}  \times {\bf B}_{ex} \right)
\end{eqnarray}
and
\begin{equation}\label{Pin2}
 {\bf P}_{in,2}(t) = \epsilon_0 \int d \x \,  \left(     ( \nabla \cdot  {\bf A}  ) {\bf E}_{ex} -
 \nabla \Phi  \times {\bf B}_{ex}  \right).
\end{equation}
In the above we additionally used Maxwell equation $ \nabla \times {\bf E}_{ex} = - \partial_t {\bf B}_{ex}$.

\subsection{Calculation of $ {\bf P}_{in,1}$}

We now concentrate our attention on $ {\bf P}_{in,1}$ given by Eq.~(\ref{Pin1}).
As ${\bf A}$ field is initially zero thus we have
\begin{equation} \label{Pin1In}
{\bf P}_{in,1}(t_i) = 0.
\end{equation}
We now calculate ${\bf P}_{in,1}(t_f)$.
We have
\begin{eqnarray} \nonumber
&& 
{\bf P}_{in,1}(t_f) =
- \epsilon_0 \int d \x \, \left( 
- {\bf A} \times   \partial_{t_f}{\bf B}_{ex} 
+ \partial_{t_f} {\bf A}  \times {\bf B}_{ex} \right)
\\ \nonumber
\\  \nonumber
&&
=
 \int d\K \, 
\int d \x' \int_{-\infty}^{t_f} dt' \,   
   {\bf J}(\x',t')   \sin(kc(t_f-t')) \times
\left(  - i e^{-i ck t_f + i \K\x'} {\bf B}_-(\K) +i  e^{i ck t_f + i \K\x'} {\bf B}_+(\K)     \right) 
\\ \nonumber
\\ \nonumber
&& 
-  \int d\K \, 
\int d \x'  \int_{-\infty}^{t_f}  dt' \,   
   {\bf J}(\x',t')  \cos(kc(t_f-t')) \times
\left(   e^{-i ck t_f + i \K\x'} {\bf B}_-(\K) + e^{i ck t_f + i \K\x'} {\bf B}_+(\K)     \right) 
\\ \nonumber
\\ \label{Pin1Fi}
&&
= - \int d \x'  \int_{-\infty}^{t_f}  dt' \,   {\bf J}(\x',t') \times   {\bf B}_{ex}(\x',t')
\end{eqnarray}
where we introduced again $t_f$ as the upper limit of the integral.
In the above we used Eqs.~(\ref{Ech}), (\ref{A}),  (\ref{decB}) and (\ref{Gf}).

\subsection{Calculation of $ {\bf P}_{in,2}$}

We now move to ${\bf P}_{in,2}(t)$ given by Eq.~(\ref{Pin2}).
We define 
\begin{equation}\label{Pin21}
{\bf P}_{in,2} = {\bf P}_{in,2,1} +  {\bf P}_{in,2,2} 
\ \ \ \ \ 
{\bf P}_{in,2,1} =  \epsilon_0 \int d \x \,  (  \nabla \cdot {\bf A}) {\bf E}_{ex}
\ \ \ \ \ 
 {\bf P}_{in,2,2} = 
 - \epsilon_0 \int d \x \,  
 \nabla \Phi  \times {\bf B}_{ex}. 
\end{equation}

\subsubsection{Calculation of  ${\bf P}_{in,2,1}$ }

As ${\bf A}$ is zero for $t=t_i$ thus
\begin{equation}\label{Pin21In}
{\bf P}_{in,2,1}(t_i) = 0.
\end{equation}
We need to calculate
\begin{equation} \label{Pin21final}
{\bf P}_{in,2,1}(t_f)
=  \epsilon_0 \int d \x \,  {\bf E}_{ex} (\x,t)   \frac{\mu_0}{4\pi} \int d \x' \int_{t_i}^{t_f} dt' \, 
\nabla_\x \cdot \left(  G(\x-\x',t_f-t') {\bf J}(\x',t') \right)
\end{equation}
where we used Eq.~(\ref{A}) and introduced $t_f$ and $t_i$ instead of $\infty$ and $-\infty$ respectively (here we need to remember that ${\bf J}(\x',t') =0$ for $t'< 0$).
Now we use
\begin{eqnarray} \nonumber
&& \nabla_\x \cdot \left(  G(\x-\x',t_f-t') {\bf J}(\x',t') \right) 
= \sum_i  \left( \partial_{x_i}  G(\x-\x',t_f-t') \right) J_i(\x',t')
\\ \nonumber
&&
=  \sum_i  \left(  - \partial_{x_i'}  G(\x-\x',t_f-t') \right) J_i(\x',t')
= - \sum_i   \partial_{x_i'}  \left( G(\x-\x',t_f-t') J_i(\x',t') \right)
+ \sum_i  G(\x-\x',t_f-t')  \partial_{x_i'} J_i(\x',t')
\\ \label{rr1}
&& =  -    \nabla_{\x'} \cdot  \left( G(\x-\x',t_f-t') {\bf J}(\x',t') \right)
+  G(\x-\x',t_f-t')  \nabla_{\x'} \cdot {\bf J}(\x',t')
\end{eqnarray}
We have
\begin{equation} \label{rr2}
  \epsilon_0 \int d \x \,  {\bf E}_{ex} (\x,t)   \frac{\mu_0}{4\pi} \int d \x' \int_{t_i}^{t_f} dt' \, 
\nabla_{\x'} \cdot \left(  G(\x-\x',t_f-t') {\bf J}(\x',t') \right) =0.
\end{equation}
To see the above we need to take the volume as having any shape on which boundary ${\bf J}(\x',t') =0 $ which may be easily obtained.
Therefore from Eqs.~(\ref{Pin21final}), (\ref{rr1}) and (\ref{rr2}) we obtain
\begin{eqnarray} \nonumber
&&  {\bf P}_{in,2,1}(t_f) =  \epsilon_0 \int d \x \,  {\bf E}_{ex} (\x,t_f)   \frac{\mu_0}{4\pi} \int d \x' \int_{t_i}^{t_f} dt' \,  G(\x-\x',t_f-t') \nabla_{\x'} \cdot {\bf J}(\x',t')
\\ \nonumber
&& = 
  - \frac{1}{4\pi c^2} \int d \x \,  {\bf E}_{ex}(\x,t_f)     \int d \x' \int_{t_i}^{t_f} dt' \,  G(\x-\x',t_f-t') \partial_{t'} \rho (\x',t')
\\  \nonumber
&& =  - \frac{1}{4\pi c^2} \int d \x \,  {\bf E}_{ex} (\x,t_f)    \int d \x' \int_{t_i}^{t_f} dt' \,  \partial_{t'} ( G(\x-\x',t_f-t')  \rho (\x',t')) - \left( \partial_{t'} G(\x-\x',t_f-t') \right)  \rho (\x',t') 
\\ \label{Pin21Fi}
&& = {\bf P}_{in,2,1,1}(t_f) + {\bf P}_{in,2,1,2}(t_f)
\end{eqnarray}
where we used  the continuity equation 
\begin{eqnarray*}
\nabla_{\x'} \cdot {\bf J}(\x',t') +\partial_{t'} \rho (\x',t')=0.
\end{eqnarray*}
We continue
\begin{eqnarray*}
{\bf P}_{in,2,1,1}(t_f) &=& -  \frac{1}{4\pi c^2} \int d \x \,  {\bf E}_{ex}(\x,t_f)     \int d \x' \int_{t_i}^{t_f}  dt' \,  \partial_{t'} (G(\x-\x',t_f-t')  \rho (\x',t'))
\\
&=&  
-  \frac{1}{4\pi c^2} \int d \x \,  {\bf E}_{ex}(\x,t_f)     \int d \x' 
\,  \left( G(\x-\x',0) \rho(\x',t_f) - G(\x-\x',t_f-t_i)  \rho (\x',t_i) \right)
\\
&=& -  \frac{1}{4\pi c^2} \int d \x \,  {\bf E}_{ex}(\x,t_f)     \int d \x'  \, \delta(\x-\x')  \rho(\x',t_f)  + 
  \frac{1}{4\pi c^2} \int d \x \,  {\bf E}_{ex}(\x,t_f)  \int d \x' \, G(\x-\x',t_f-t_i)  \rho (\x',t_i)
\\
&=& -  \frac{1}{4\pi c^2} \int d \x \,  {\bf E}_{ex}(\x,t_f)    \rho(\x,t_f)  + 
  \frac{1}{4\pi c^2} \int d \x \,  {\bf E}_{ex}(\x,t_f)  \int d \x' \, G(\x-\x',t_f-t_i)  \rho (\x',t_i)
\end{eqnarray*}
where  we used $ G(\x-\x',0) = \delta(\x-\x') $.
As $  {\bf E}_{ex}(\x,t_f)    \rho(\x,t_f) = 0  $ thus we have
\begin{eqnarray}\nonumber
&& {\bf P}_{in,2,1,1}(t_f)  =
  \frac{1}{4\pi c^2} \int d \x \,  {\bf E}_{ex}(\x,t_f)     \int d \x'  \,   G(\x-\x',t_f-t_i)  \rho (\x',t_i)
=  \frac{q}{4\pi c^2} \int d \x \,  {\bf E}_{ex}(\x,t_f)      \frac{\delta \left( |\x| - c(t_f - t_i)   \right)}{|\x|} 
\\ \label{Pin211final}
&& = \frac{q}{4\pi c^2}  \frac{3}{4} cT  f_0 {\bf e}_x \frac{1}{c(t_f-t_i)} \frac{c(t_f-t_i)}{c \sqrt{(t_f-t_i)^2 - t_f^2}} 2\pi c \sqrt{(t_f-t_i)^2 - t_f^2}
=   \frac{3}{8} q T  f_0   {\bf e}_x
\end{eqnarray}
where we used Eq.~(\ref{Gdef}) and $\rho(\x',t_i) = q \delta(\x')$ and we used $t_f \gg T$.
\\
\\
From Eq.~(\ref{Pin21Fi}) we have
\begin{eqnarray}  \nonumber
&& {\bf P}_{in,2,1,2}(t_f) = 
  \frac{1}{4\pi c^2} \int d \x \,  {\bf E}_{ex} (\x,t_f)    \int d \x' \int_{t_i}^{t_f} dt' \,  \left( \partial_{t'} G(\x-\x',t_f-t') \right)  \rho (\x',t') 
\\ \nonumber
&& 
= -  \frac{1}{4\pi c^2}\int d \x' \int_{t_i}^{t_f} dt' \, \rho (\x',t') \int d \x \,  {\bf E}_{ex} (\x,t_f)   \partial_{t_f}  G(\x-\x',t_f-t')   
\\ \label{Pin212final}
&& = - \frac{1}{2}   \int d \x' \int_{t_i}^{t_f} dt' \, \rho (\x',t')  \left( {\bf E}_{ex}(\x',2t_f-t') + {\bf E}_{ex}(\x',t')  \right)
= 
- \frac{1}{2}   \int d \x' \int_{t_i}^{t_f} dt' \, \rho (\x',t')  {\bf E}_{ex}(\x',t')  
\end{eqnarray} 
where we used $ \partial_{t'} G(\x-\x',t_f-t')  = - \partial_{t_f}  G(\x-\x',t_f-t')   $, Eq.~(\ref{WAZNE2}) and additionally the fact that
$  \rho (\x',t')  {\bf E}_{ex}(\x',2t_f-t')  = 0  $ for $ t_i <  t' < t_f$.

% t_0 jest na tyle duze ze sygnal wyslany z pradu/ladunku w czasie t0 nie ma szans dotrzec do Eex.
% Wiec ta pierwsza czesc jest zero.  

\subsubsection{Calculation of  ${\bf P}_{in,2,2}$ }

Now we turn to ${\bf P}_{in,2,2}$ given by Eq.~(\ref{Pin21}).
Its value at $t=t_i$ is equal to
\begin{eqnarray} \nonumber
&& 
{\bf P}_{in,2,2}(t_i) =  - \epsilon_0 \int d \x \,  \nabla \Phi(\x,t_i)  \times {\bf B}_{ex}(\x,t_i) 
=  - \epsilon_0 \int d \x \,    \frac{q  ( x {\bf e}_z - z {\bf e}_x )}{4\pi \epsilon_0  (x^2 + y^2 + z^2)^{3/2}} 
f(z-ct_i)
\\ \nonumber
\\ \nonumber
&& = 
 - \int  dx dy \int_{-c |t_i|- 3T/4}^{-c |t_i|}  d z \,    \frac{q  ( x {\bf e}_z - z {\bf e}_x )}{4\pi   (x^2 + y^2 + z^2)^{3/2}} f_0
= 
 \int  dx dy \int_{-c |t_i|- 3T/4}^{-c |t_i|}  d z \,    \frac{q   z {\bf e}_x }{4\pi   (x^2 + y^2 + z^2)^{3/2}} f_0
\\ \nonumber
\\ \label{Pin22In}
&&
=
 \frac{q  {\bf e}_x  f_0 }{4\pi  } \int_{-c |t_i|- 3T/4}^{-c |t_i|}  d z 
    \int  d\tilde x d \tilde y \, \frac{1}{(\tilde x^2 + \tilde y^2 +1)^{3/2}}
=  \frac{3}{8 } q   f_0 T{\bf e}_x 
\end{eqnarray}
Now we turn our attention to 
 \begin{eqnarray}\nonumber 
  {\bf P}_{in,2,2}(t_f) 
= - \epsilon_0 \int d \x \,  \nabla \times   \left(\Phi {\bf B}_{ex} \right)(t_f)
+ \epsilon_0 \int d \x \, 
  \Phi     \nabla  \times {\bf B}_{ex} 
\end{eqnarray}
where we used $(\nabla \Phi)  \times {\bf B} = - \nabla \times ( \Phi {\bf B}) + \Phi \nabla \times {\bf B}$.
We find that 
\begin{eqnarray*}
 \epsilon_0 \int d \x \,  \nabla \times   \left(\Phi {\bf B}_{ex} \right)(t_f) = 0. 
\end{eqnarray*}
This can be seen by taking the volume as a cube of length $L$ larger than $2c t_f$. The center of this cube is in the initial position of the static particle.
The  normal vectors to the surface of this cube are equal to $\pm {\bf e}_{x}, \pm {\bf e}_{y}, \pm {\bf e}_{z} $. Than the surface integral 
calculated on the wall with normal vector $ {\bf e}_x$ is nonzero but is exactly canceled by the integral on the opposite wall (with normal vector $-{\bf e}_x$).
As a result we have
\begin{eqnarray} \nonumber
&&  {\bf P}_{in,2,2}(t_f)  =  \epsilon_0 \int d \x \, 
  \Phi(\x,t_f)     \nabla  \times {\bf B}_{ex}(\x,t_f)
=  \epsilon_0 \int d \x \, 
  \Phi(\x,t_f)    \frac{1}{c^2} \partial_{t_f} {\bf E}_{ex}(\x,t_f)
\\ \nonumber
&& =
\int d\K \, 
 \left( - i e^{-i ck t_f + i \K\x'} {\bf E}_-(\K) + i e^{i ck t_f + i \K\x'} {\bf E}_+(\K)     \right) 
 \int d \x' \int_{-\infty}^{t_f}  dt' \,   
 \rho (\x',t')   \sin(kc(t_f-t'))
\\ \label{Pin22Fi}
&& = \frac{1}{2}   \int d \x' \int_{-\infty}^{t_f} dt' \, \rho (\x',t')  \left(  {\bf E}_{ex}(\x',2t_f-t') - {\bf E}_{ex}(\x',t')  \right)
=  - \frac{1}{2}   \int d \x' \int_{-\infty}^{t_f} dt' \, \rho (\x',t')  {\bf E}_{ex}(\x',t')  
\end{eqnarray}
where in the above we used Maxwell equation $\nabla  \times {\bf B}_{ex}(\x,t_f) =  \frac{1}{c^2}\partial_{t_f} {\bf E}_{ex}(\x,t_f)$
and Eqs.~(\ref{Phi}), (\ref{decE}) and (\ref{Gf}). In addition we again made use of the fact that 
$  \rho (\x',t')  {\bf E}_{ex}(\x',2t_f-t')  = 0  $ for $   t' < t_f$.

\subsubsection{Formula for $ {\bf P}_{in,2}$ }

From Eqs.~(\ref{Pin21In}), (\ref{Pin21Fi}),  (\ref{Pin211final}),  (\ref{Pin212final}), (\ref{Pin22In}) and (\ref{Pin22Fi}) we find that
\begin{equation} \label{Pin2wynik}
 {\bf P}_{in,2}(t_i) =  \frac{3}{8} q T  f_0   {\bf e}_x
\ \ \ \ \ \ 
 {\bf P}_{in,2}(t_f) = \frac{3}{8} q T  f_0   {\bf e}_x  - \int d \x' \int_{-\infty}^{\infty} d t'  \,  {\bf E}_{ex}(\x',t') \rho(\x',t').
\end{equation}
where in the above we extended the region of integration over $t'$ as in this added part $ {\bf E}_{ex}(\x',t') \rho(\x',t') =0$.

\subsection{Formulas for  ${\bf P}_{in}$ }

As a result from %Eqs.~(\ref{PinitialNN}),  (\ref{PfinalNN}),
Eqs.~(\ref{Pindivision}), (\ref{Pin1In}), (\ref{Pin1Fi}) and  (\ref{Pin2wynik}) we find that
\begin{eqnarray}\nonumber
&& {\bf P}_{in}(t_i) =    \frac{3}{8} q T  f_0   {\bf e}_x
\\  \label{PinWYNIK}
\\ \nonumber
&& {\bf P}_{in}(t_f) =    \frac{3}{8} q T  f_0   {\bf e}_x    - q \int_{-\infty}^{\infty}  dt  \,   \left( {\bf E}_{ex}(\x(t),t) + {\bf v}(t) \times {\bf B}_{ex}(\x(t),t) \right) 
\end{eqnarray}
where we used  ${\bf J}(\x,t) = q {\bf v}(t) \delta(\x-\x(t))$, $\rho(\x,t) = q \delta(\x-\x(t))  $.

\subsection{Formulas for initial and final momentum of the system}

As a result from Eqs.~(\ref{PinitialNN}),  (\ref{PfinalNN}) and (\ref{PinWYNIK}) we find
\begin{eqnarray}\nonumber 
&& {\bf P}_i =   {\bf P}_{ex} +  \frac{3}{8} q T  f_0   {\bf e}_x
\\ \label{PWynik}
\\ \nonumber
&& {\bf P}_f =     {\bf P}_{ex}     + \widetilde {\bf P}_{rad}  + \gamma_f m {\bf v}_f   +  \frac{3}{8} q T  f_0   {\bf e}_x    - q \int_{-\infty}^{\infty}  dt  \,   \left( {\bf E}_{ex}(\x(t),t) + {\bf v}(t) \times {\bf B}_{ex}(\x(t),t) \right) 
\end{eqnarray}

\end{widetext}


\begin{thebibliography}{99}
\bibitem{dopiska4}
Here we assume that when no external forces act on the charged particle for very long time it moves with constant velocity.
If this were not the case the particle would accelerate and lose energy.
\bibitem{jeszczeDopiska}
%We consider the theories in which causal particle trajectories {\it exists}
%for all kinds of pulses (even if the trajectory is unphysical).
We consider the theories in which renormalization procedures present are equivalent to the renormalization procedure defined in this paper, when used for static particle resting for infinitely long time.
However this seems to be obvious, as it is seem almost impossible 
to have another renormalization procedure in the case of static particle, as the one described in the present paper.
\bibitem{dopiskaFeynman}
The fact that Lorentz model is inconsistent was noted in \cite{Feynman}
where Feynman writes
``Non-electric forces are required to hold together the charge distribution, according to Poincare, for to neglect such forces is to violate the relativistic relation between mass and energy. A composite system of this kind would possess an infinite number of internal degrees of freedom of oscillation. No consistent model has been found for the Lorentz electron in either classical or quantum mechanics."
``Briefly, Lorentz attempts to propose a physical mechanism behind the radiation reaction, but arrives at a mathematically incomplete expression for this force."
\end{thebibliography}
\end{document}